\documentclass{emulateapj}
\bibpunct{(}{)}{;}{a}{}{,}

\newcommand{\rh}{\varrho}
\newcommand{\vunit}{\mbox{m}\,\mbox{s}^{-1}}
\newcommand{\dunit}{\mbox{m}^2\,\mbox{s}^{-1}}
\renewcommand{\vec}[1]{\mbox{\boldmath$#1$}}

\begin{document}

\title{Flux-transport dynamos with Lorentz force feedback on differential 
  rotation and meridional flow: Saturation mechanism and torsional 
  oscillations}

\author{Matthias Rempel}

\affil{High Altitude Observatory,
       National Center for Atmospheric Research\footnote{The National
       Center for Atmospheric Research is sponsored by the National
       Science Foundation} , 
       P.O. Box 3000, Boulder, Colorado 80307, USA
      }

\email{rempel@hao.ucar.edu}

\shorttitle{Dynamos with Lorentz force feedback}
\shortauthors{M. Rempel}

\begin{abstract}
In this paper we discuss a dynamic flux-transport dynamo model that includes
the feedback of the induced magnetic field on differential rotation and 
meridional flow. We consider two different approaches for the feedback: 
mean field Lorentz force and quenching of transport coefficients such as 
turbulent viscosity and heat conductivity. We find that even strong feedback on
the meridional flow does not change the character of the flux-transport dynamo
significantly; however it leads to a significant reduction of differential
rotation. To a large degree independent from the dynamo parameters, the
saturation takes place when the toroidal field at the base of the convection
zone reaches between $1.2$ an $1.5\,\mbox{T}$, the energy converted into
magnetic energy corresponds to about $0.1\%$ to $0.2\%$ of the solar 
luminosity. 
The torsional oscillations produced through Lorentz force feedback on
differential rotation show a dominant poleward propagating branch with the
correct phase relation to the magnetic cycle. We show that incorporating 
enhanced surface cooling of the active region belt (as proposed by Spruit) 
leads to an equatorward propagating branch in good agreement with 
observations.   
\end{abstract}

\keywords{Sun: interior --- rotation --- magnetic field --- dynamo}

\section{Introduction}
Flux-transport dynamos have proven to be successful
for modeling the evolution of the large scale solar magnetic field
\citep{Wang:Sheeley:1991,Durney:1995,Choudhuri:etal:1995,
Dikpati:Charbonneau:1999,Kueker:etal:2001,Dikpati:Gilman:2001:dynamo,
Dikpati:etal:2004,Dikpati:2005,Dikpati:etal:2006:pred}. 
In a flux-transport dynamo the 
equatorward propagation of the magnetic activity belt (butterfly diagram)
is a consequence of the equatorward transport of magnetic field at the
base of the convection zone by the meridional flow. 

However, all studies so far have addressed the transport of magnetic field 
by the
meridional circulation in a purely kinematic regime. The toroidal field
strength at the base of the solar convection zone inferred from studies
of rising magnetic flux tubes \citep{Choudhuri:Gilman:1987,Fan:etal:1993,
Schuessler:etal:1994,Caligari:etal:1995,Caligari:etal:1998}
is around $10\,\mbox{T}$ ($100\,\mbox{kG}$) and thus orders of magnitude
larger than the equipartition field strength estimated from a meridional
flow velocity of a few $\vunit$. Therefore it is crucial for flux-transport
dynamos to include the feedback of the Lorentz force on the meridional flow. 

In order to be able to address this question it is necessary to incorporate
a model for the solar differential rotation and meridional flow into a
dynamo model and allow for the feedback of the Lorentz force on differential
rotation and meridional flow. Differential rotation and meridional flow
have been addressed in the past mainly through two approaches: 3D full
spherical shell simulations \citep{Glatzmaier:Gilman:1982,Gilman:Miller:1986,
Miesch:etal:2000,Brun:Toomre:2002} and axisymmetric mean field models
\citep{Kitchatinov:Ruediger:1993,Kitchatinov:Ruediger:1995,Ruediger:etal:1998,
Kueker:Stix:2001}. While the 3D simulations have trouble reproducing a
consistent large scale meridional flow pattern (poleward in the upper half 
of the convection zone), as it is inferred by helioseismology 
\citep{Braun:Fan:1998,Haber:etal:2002,Zhao:Kosovichev:2004}, such a 
flow is a common feature in most of the mean field models.

In this paper we build upon the differential rotation model presented in 
\citet{Rempel:2005} and combine it with the axisymmetric mean field dynamo
equations. The magnetic field is allowed to feed back on differential
rotation and meridional flow through the mean field Lorentz force and 
quenching of turbulent viscosity and heat conductivity. We have addressed 
already in a previous paper \citep{Rempel:2005:transport} the feedback of 
magnetic
field on the meridional flow by imposing a static toroidal magnetic
and including magnetic tension and quenching of turbulent viscosity
in the differential rotation model. We found that a significant feedback
can be expected if the toroidal magnetic field strength is around
$3\,\mbox{T}$ or above equipartition. In this paper we do not impose 
a toroidal field, but rather solve the induction equation to obtain a 
time dependent magnetic field. Going beyond \citet{Rempel:2005:transport}, 
we also include the feedback on differential rotation leading to a solar 
cycle variation of the rotation rate, which is known as torsional oscillations 
\citep{Howard:Labonte:1980,Toomre:etal:2000,Howe:etal:2000:solphys,
Antia:Basu:2001,Vorontsov:etal:2002,Howe:etal:2005}.

A very similar approach was taken before by
\citet{Brandenburg:etal:1990,Brandenburg:etal:1991,Brandenburg:etal:1992}.
In their model they were solving a mean field differential rotation model
parallel to the dynamo equation to obtain a 'dynamic' dynamo. The main
difference in our approach is that we focus on flux-transport dynamos,
whereas their work described mainly $\alpha\Omega$-dynamos with the advection
of field by a meridional flow playing only a secondary role. Given recent
developments in solar dynamo theory, showing that flux transport dynamos are
very successful in reproducing most of the observed features
\citep{Dikpati:2005} it is important to evaluate to which
degree these dynamos change operation if dynamic feedback is considered.

Lorentz force feedback has been considered in mean field dynamo models in 
various levels of sophistication:

\citet{Kueker:etal:1999} used a model including both macro (mean field Lorentz 
force) and micro (quenching of $\Lambda$-effect) feedback to evaluate to
which extent grand minima can be produced through feedback on differential
rotation.

\citet{Covas:etal:2000,Covas:etal:2004,Covas:etal:2005} considered feedback on 
differential rotation in a classical $\alpha\Omega$-dynamo model solving a 
simplified momentum equation including the mean field Lorentz force and a 
diffusive 
relaxation term for the longitudinal flow velocity perturbation. They were able
to reproduce the basic features of the observed solar torsional oscillation
pattern (equatorward and poleward propagating branch). A similar approach has
been taken before by \citet{Moss:Brooke:2000,Tobias:1996:nonlin,
Yoshimura:1981}.

In contrast to this the models of \citet{Schuessler:1979:dynamo,
Brandenburg:etal:1990,Brandenburg:etal:1991,Brandenburg:etal:1992,
Jennings:1993,Moss:etal:1995,Muhli:etal:1995} incorporate the full 
momentum equation, allowing also for magnetically driven meridional
motions. The main focus of their work was on understanding the non-linear
saturation of the dynamo.

The consideration of the macroscopic mean field Lorentz force, common for all
models listed above, is also known in the literature as 'Malkus-Proctor-effect'
\citep{Malkus:Proctor:1975}.

A different approach has been used by
\citet{Kitchatinov:Pipin:1998} and \citet{Kitchatinov:etal:1999}, who 
considered feedback through quenching of the $\Lambda$-effect (turbulent 
angular momentum transport driving differential rotation). Their work
focused on understanding torsional oscillations as well as the possibility
of producing grand activity cycles through this type of feedback.

This paper is organized as follows: In section \ref{model} we explain the 
physics included in the non-kinematic dynamo model. Section \ref{dynamicdynamo}
shows the results of the non-kinematic dynamo runs including a detailed 
analysis of the energy flows within the model.  Section \ref{torsional}
focuses on the properties of the torsional oscillations produced by the
model and compares them to results obtained by helioseismology. Section
\ref{parameter} discusses the choices we make for various parameters of
the mean field model and their impact on the solutions presented here. 
In section \ref{discussion} we summarize the main results of this 
investigation and discuss them in the context of solar dynamo models.

\section{Model}
\label{model}
In this paper we utilize the mean field differential rotation and meridional 
circulation model of \citet{Rempel:2005} and couple it with the axisymmetric 
mean field dynamo equations. The computed differential rotation and meridional 
flow are used to advance the magnetic field by using a Babcock-Leighton flux
transport dynamo model. The computed magnetic field is allowed to feed back
on differential rotation and meridional flow through the Lorentz force and
quenching of turbulent viscosity and thermal heat conductivity. 
We are solving the axisymmetric MHD equations including 
parameterizations of processes on the (unresolved) convective scale
(mean field approach). We introduce in the induction equation the vector
potential for the poloidal field to satisfy the constraint 
$\vec{\nabla}\cdot\vec{B}=0$. 
\begin{eqnarray}
   \frac{\partial \rh_1}{\partial t} &=& -\frac{1}{r^2}
      \frac{\partial}{\partial r}\left(r^2 v_r\rh_0\right)
      -\frac{1}{r\sin\theta}\frac{\partial}{\partial \theta}
      \left(\sin\theta v_{\theta}\rh_0\right)\label{dens}\\
   \frac{\partial v_r}{\partial t} &=& -v_r\frac{\partial v_r}{\partial r}
      -\frac{v_{\theta}}{r}\frac{\partial v_r}{\partial \theta}
      +\frac{v_{\theta}^2}{r}
      -\frac{\partial}{\partial r} \frac{p_{\rm tot}}{\rh_0}+
      \frac{p_{\rm mag}}{\gamma p_0}g+\frac{s_1}{\gamma}g  \nonumber \\
      && +\left(2\Omega_0\Omega_1+\Omega_1^2\right)r\sin^2\theta 
      +\frac{1}{\rh_0}\left(F_r^{\nu}+F_r^{B}\right)
      \label{vrad}\\
   \frac{\partial v_{\theta}}{\partial t} &=& 
      -v_r\frac{\partial v_{\theta}}{\partial r}
      -\frac{v_{\theta}}{r}\frac{\partial v_{\theta}}{\partial \theta}
      -\frac{v_r v_{\theta}}{r}
      -\frac{1}{r}\frac{\partial}{\partial\theta}\frac{p_{\rm tot}}{\rh_0}
      \nonumber \\
      && +\left(2\Omega_0\Omega_1+\Omega_1^2\right)r\sin\theta\cos\theta 
      +\frac{1}{\rh_0}\left(F^{\nu}_{\theta}+F^{B}_{\theta}\right)
      \label{vthe} \\
   \frac{\partial \Omega_1}{\partial t} &=& 
      -\frac{v_r}{r^2}\frac{\partial}{\partial r}\left[r^2(\Omega_0+\Omega_1)
      \right]\nonumber\\
      &&-\frac{v_{\theta}}{r \sin^2\theta}\frac{\partial}{\partial \theta}
      \left[\sin^2\theta(\Omega_0+\Omega_1)\right]\nonumber\\
      &&+\frac{1}{\rh_0 r\sin\theta}\left(F^{\nu}_{\phi}+F^{B}_{\phi}\right)
      \label{omeg}\\
   \frac{\partial s_1}{\partial t} &=& -v_r\frac{\partial s_1}{\partial r}
      -\frac{v_{\theta}}{r}\frac{\partial s_1}{\partial \theta}
      +v_r\frac{\gamma\delta}{H_p}+\frac{\gamma-1}{p_0}Q\nonumber\\
      &&+\frac{1}{\rh_0 T_0}\vec{\nabla}\cdot 
      (\kappa_t\rh_0 T_0\vec{\nabla} s_1)
      +\frac{\gamma-1}{p_0}\eta_t(\vec{\nabla}\times\vec{B})^2
      \label{entr}\\
   \frac{\partial B_{\Phi}}{\partial t} &=& -\frac{1}{r}
      \frac{\partial}{\partial r}\left(r v_r B_{\Phi}\right)-\frac{1}{r}
      \frac{\partial}{\partial \theta}\left(v_{\theta} B_{\Phi}\right)
      +r\sin\theta B_r\frac{\partial \Omega_1}{\partial r}\nonumber\\
      &&+\sin\theta B_{\theta}\frac{\partial \Omega_1}{\partial\theta}
      +\eta_t\left(\Delta-\frac{1}{(r\sin\theta)^2}\right)B_{\Phi}\nonumber\\
      &&+\frac{1}{r}\frac{\partial \eta_t}{\partial r}
      \frac{\partial}{\partial r}
      \left(r B_{\Phi}\right)+\frac{1}{r^2}\frac{\partial \eta_t}
      {\partial \theta}\frac{1}{\sin\theta}\frac{\partial}{\partial \theta}
      \left(\sin\theta B_{\Phi}\right)\label{bphi} \\
   \frac{\partial A}{\partial t} &=&  -\frac{v_r}{r}\frac{\partial}{\partial r}
      \left(r A\right)-\frac{v_{\theta}}{r\sin\theta}\frac{\partial}
      {\partial \theta}\left(\sin\theta A\right)\nonumber\\
      &&+\eta_t\left(\Delta-\frac{1}{(r\sin\theta)^2}\right)A
      +S\left(r,\theta,B_{\Phi}\right)\label{apol}\;.
\end{eqnarray}
Here $\rh_0$ and $p_0$ denote the (spherical symmetric) reference state 
stratification, $H_p=p_0/(\rh_0 g)$ the pressure scale height, $\rh_1$ and 
$p_1$ perturbations around the reference state caused by differential rotation 
and meridional flow. Since these perturbations are small compared to 
the reference state values, the equations are linearized assuming
$\rh_1\ll \rh_0$ and $p_1\ll p_0$. $\Omega_0$ denotes the
rotation rate of the core, $\Omega_1$ the differential rotation with respect
to the core in the convection zone. The quantity 
$s_1=p_1/p_0-\gamma\rh_1/\rh_0$
denotes the dimensionless entropy perturbation (normalized by the heat capacity
$c_v$). For the reference state we use an adiabatic polytrope assuming a 
gravity varying $\sim r^{-2}$; however, small perturbations from adiabaticity
are considered in the entropy equation through the third term 
$\sim\delta=\nabla-\nabla_{\rm ad}$. 
The quantity $p_{\rm mag}$ denotes the magnetic pressure, 
$p_{\rm tot}=p_1+p_{\rm mag}$ the total pressure. The buoyancy term in
Eq. (\ref{vrad}) has been written in a way to separate the magnetic buoyancy
from non-magnetic buoyancy, assuming $\vert\nabla-\nabla_{\rm ad}\vert\ll 1$.
Since the most unstable modes driven by magnetic buoyancy are typically 
non-axisymmetric, the description of magnetic buoyancy in our axisymmetric
model is not necessarily very realistic. Therefore we will discuss later also
simulations with magnetic buoyancy switched off by ignoring the term 
$\sim p_{\rm mag}$ in Eq. (\ref{vrad}).

The quantity $\vec{F}^{B}=1/\mu_0\vec{\nabla}\cdot\left(\vec{B}\vec{B}\right)$
denotes magnetic tension,  where
the poloidal magnetic field follows from the vector potential $A$ used in
Eq. (\ref{apol}) through
\begin{eqnarray}
  B_r&=&\frac{1}{r\sin\theta}\frac{\partial}{\partial\theta}
  \left(\sin\theta A\right)\\
  B_{\theta}&=&-\frac{1}{r}\frac{\partial}{\partial r}\left(r A\right)\;.
\end{eqnarray}
For computing the Lorentz force we consider here only the magnetic mean field
contribution. Formally an additional contribution to the stress tensor 
$\sim \langle\vec{B}^{\prime}\vec{B}^{\prime}\rangle$ 
caused by the turbulent magnetic 
field exists; however the contribution of these terms in detail is not well
understood. 
We emphasize that considering only the mean field Lorentz force leads to a 
model that is energetically consistent in the way that the energy extracted
from differential rotation and meridional flow is flowing into the reservoir 
of magnetic energy via the induction equation. Considering additional 
Lorentz force terms in the
momentum equation requires that these terms only redistribute momentum
(otherwise additional terms in the induction equation would be required as 
well to be energetically consistent). This effect referred to as 
visco-elasticity by some authors \citep{Longcope:etal:2003} could 
parametrized to some extent as an additional source of viscous stress. Another
possible feedback of these terms is quenching of turbulent transport processes
such as turbulent angular momentum transport, viscosity and heat conductivity,
which we will consider later in the form:
\begin{eqnarray}
  \nu_t&=&\frac{\nu_t^0}{1+\left(B/B_{\rm eq}\right)^2}\label{eta_quench}\\
  \kappa_t&=&\frac{\kappa_t^0}{1+\left(B/B_{\rm eq}\right)^2}
  \label{kappa_quench}
\end{eqnarray}

The quantity $\vec{F}^{\nu}$ denotes turbulent viscous stresses, including 
turbulent transport of angular momentum ($\Lambda$-effect), $Q$ considers
the associated viscous dissipation in the entropy equation. In our model the
$\Lambda$-effect is the primary driver of differential rotation and meridional
flow, while the profile of the differential rotation (the deviation from the
Taylor-Proudman state) is a consequence of a latitudinal entropy gradient.
The latitudinal entropy gradient follows in our model self consistently from
the inclusion of a subadiabatic tachocline. For further details concerning
the differential rotation model we refer to \citet{Rempel:2005} and section
\ref{diffrot}. 

Eqs. (\ref{bphi}) and (\ref{apol}) are the axisymmetric mean field dynamo
equation, including transport of magnetic field by meridional flow,
shear by differential rotation, magnetic diffusion and induction by a
Babcock-Leighton surface $\alpha$-effect, parametrized through the
poloidal source term $S\left(r,\theta,B_{\Phi}\right)$. We will discuss the
dynamo model in more detail in section \ref{dynamo}.

For the solar parameter range the system defined by Eqs. 
(\ref{dens}) - (\ref{apol}) is in the regime of highly subsonic flows, which
introduces a significant CFL time step constraint if solved explicitly. We do 
not use here the anelastic approximation as many others do; instead we use an
approach of artificially reducing the speed of sound to values that do not
impose a significant time step constraint (compared to the diffusive time step)
but still ensure that the flows remain highly subsonic. Given the fact that
the Mach number of the meridional flow in the bulk of the convection zone is
around $10^{-5}$ a decrease of the speed of sound by a factor $\sim 100$ is 
found to have no impact on the solution and is used for most of the results 
shown. Formally this is achieved by changing the equation of continuity to
\begin{equation}
   \frac{\partial \rh_1}{\partial t} = -\zeta^2\left[\frac{1}{r^2}
      \frac{\partial}{\partial r}\left(r^2 v_r\rh_0\right)
      -\frac{1}{r\sin\theta}\frac{\partial}{\partial \theta}
      \left(\sin\theta v_{\theta}\rh_0\right)\right]  
\end{equation}
with $\zeta\sim 0.01$, which reduces the speed of sound by a factor
$\zeta$. This approach is equivalent to increasing the base rotation
rate and scaling up all other variables to maintain the proper relations
between the different terms in the equations \citep{Rempel:2005}.
The modified equations are solved using a MacCormack scheme.
We emphasize that this approach is only feasible for the axisymmetric system,
where the much faster speed of rotation does not enter the CFL condition.

\begin{figure*}
  \resizebox{\hsize}{!}{\includegraphics{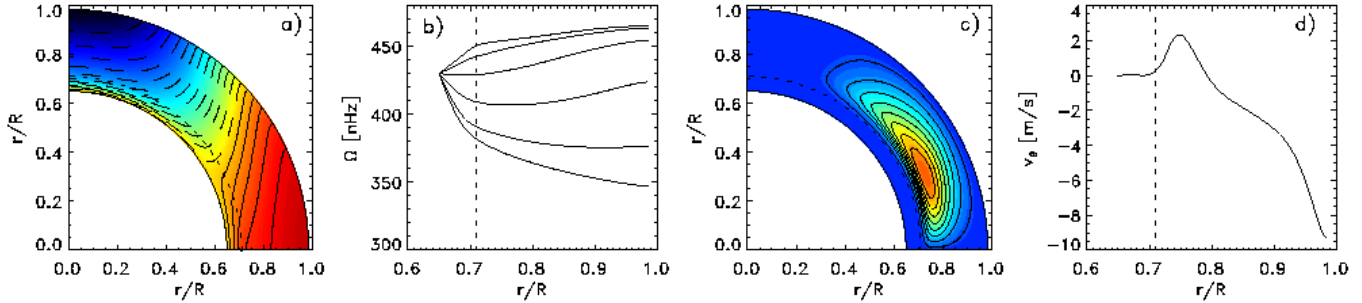}}
  \caption{Differential rotation and meridional flow of the reference model.
    a) Contours of $\Omega$, b) radial profiles of $\Omega$ for
    the latitudes (top to bottom) $0\degr$, $15\degr$, $30\degr$, 
    $45\degr$, $60\degr$,
    and $90\degr$. c) shows the stream function of the meridional flow,
    d) the flow profile at $30\degr$ latitude. The dotted line indicates the
    base of the convection zone at $r_{\rm bc}=0.71\,R_{\odot}$. Note that 
    in our model most of the radial shear is located beneath $r_{\rm bc}$,
    while the meridional flow shows only very little penetration beneath
    $r_{\rm bc}$.
  }
  \label{f1}
\end{figure*}

\subsection{Differential rotation, meridional flow reference model}
\label{diffrot}
In this paper we use a reference model that is very close to model 1
discussed in \citet{Rempel:2005}. We made the following changes in order to
obtain a meridional flow pattern that leads to a solar like dynamo period 
and a confinement of magnetic activity close to the equator:
We use a value of the parameter $n$ defining the latitudinal profile of
the $\Lambda$-effect of $3$, an amplitude of the $\Lambda$-effect of 
$\Lambda_0=1$, and a value of turbulent viscosity and heat conductivity
of $3\times 10^8\,\dunit$. $\Lambda_0$ determines primarily the amplitude
of differential rotation, while a change of $\nu_t$ and $\kappa_t$ (keeping
$\nu_t/\kappa_t$ constant) adjusts the meridional flow speed.

Fig. \ref{f1} a,b) shows the differential rotation and Fig. \ref{f1} c,d)
the meridional flow of the reference 
model. Note that in our model most of the radial shear is located beneath
the base of the convection zone $r_{\rm bc}=0.71\,R_{\odot}$. This is due
to the fact that the differential rotation is driven by the $\Lambda$-effect
within the convection and uniform rotation is imposed at the lower boundary
at $r=0.65\,R_{\odot}$. As a consequence a viscous shear layer forms between
both regions, which has the largest shear rate where the turbulent viscosity
is assumed to be small (below $r_{\rm bc}$). Also the meridional flow does not
show a significant penetration below $r_{\rm bc}$, due to the subadiabatic
stratification and the significant drop of turbulent viscosity (see 
\citet{Rempel:2005} for a detailed discussion).

\subsection{Flux-transport dynamo model}
\label{dynamo}

\begin{figure}
  \resizebox{\hsize}{!}{\includegraphics{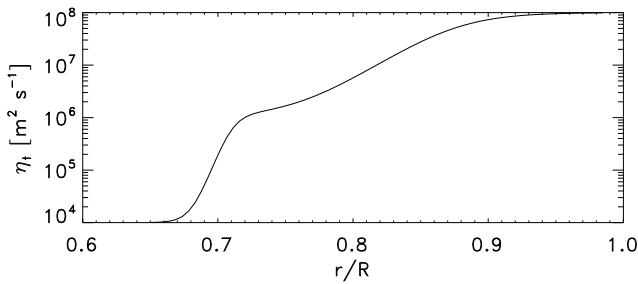}}
  \caption{Profile of $\eta_t$ used for the dynamo simulation. $\eta_t$
    drops by two orders of magnitude within the convection starting from a 
    surface value of $10^8\,\dunit$. There is an additional
    drop by one order of magnitude beneath the base of the convection zone.
  }
  \label{f2}
\end{figure}
\begin{figure}
  \resizebox{\hsize}{!}{\includegraphics{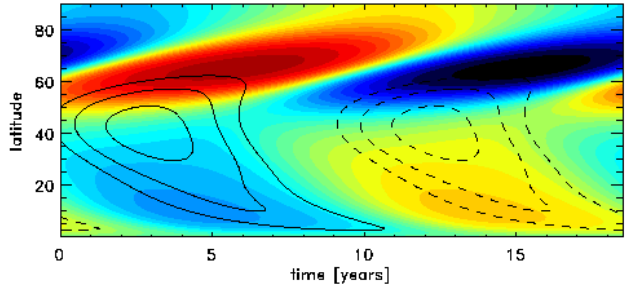}}
  \caption{Butterfly diagram of toroidal magnetic field (contour lines of
    $\bar{B}_{\Phi, {\rm bc}}$)
    and radial field close to surface $r=0.985\,R_{\odot}$
    (color shades). The maximum field strength of $B_{\Phi}$ is 
    $1.28\,\mbox{T}$, the maximum field strength of $B_r$ is $0.01\,\mbox{T}$.
  }
  \label{f3}
\end{figure}

We use a flux-transport dynamo model similar to the approach of
\citet{Dikpati:Charbonneau:1999}, except for the fact that differential
rotation and meridional flow are not prescribed, but computed by the model
described above. The additional dynamo parameters we have to specify are
the profile of the magnetic diffusivity $\eta_t$ and the functional form
of the poloidal source term $S\left(r,\theta,B_{\Phi}\right)$ in Eq. 
\ref{apol}. The turbulent magnetic diffusivity $\eta_t$ is specified as
function in radius, given by
\begin{eqnarray}
  \eta_t&=&\eta_{\rm c}+f_{\rm c}(r)\left[\eta_{\rm bc}
    -\eta_{\rm c}+f_{\rm cz}(r)\left(\eta_{\rm cz}-\eta_{\rm bc}\right)
    \right]
  \label{eta}
\end{eqnarray}
with
\begin{eqnarray}
  f_{\rm cz}(r)&=&\frac{1}{2}
    \left[1+\tanh\left(\frac{r-r_{\rm cz}}{d_{\rm cz}}\right)\right]\\
  f_{\rm c}(r)&=&\frac{1}{2}
    \left[1+\tanh\left(\frac{r-r_{\rm bc}}{d_{\rm bc}}\right)\right]\;.
\end{eqnarray}

The function $f_{\rm cz}(r)$ determines the profile within the convection zone,
while $f_{\rm c}(r)$ ensures a significant drop of $\eta_t$ beneath
$r_{\rm bc}$. $\eta_{\rm c}$ determines the core diffusivity,
$\eta_{\rm bc}$ the diffusivity at the base of the convection zone, and
 $\eta_{\rm cz}$ the diffusivity in the upper half of the convection zone.
For this discussion we use a profile with the parameters
$\eta_{\rm c}=10^5\,\dunit$, $\eta_{\rm bc}=10^6\,\dunit$,
$\eta_{\rm cz}=10^8\,\dunit$,
$r_{\rm cz}=0.875\,R_{\odot}$,
$d_{\rm cz}=0.05\,R_{\odot}$, $r_{\rm bc}=0.71\,R_{\odot}$, 
$d_{\rm bc}=0.0125\,R_{\odot}$. The profile is shown in Fig. \ref{f2}. 
We use here for reasons of numerical stability a value of $\eta_{\rm c}$ that
is significantly larger than molecular resistivity. However, the influence of 
$\eta_{\rm c}$ on the solution is found to be very weak. 

Our model uses a non-local Babcock-Leighton $\alpha$-effect 
\citep{Babcock:1961,Leighton:1969} in which the
source term $S\left(r,\theta,B_{\Phi}\right)$ at the surface is dependent on 
the toroidal field strength at the base of the convection zone averaged over
the interval $[0.71\,R_{\odot}, 0.76\,R_{\odot}]$, 
$\bar{B}_{\Phi, {\rm bc}}$. The functional form of $S$ is then
\begin{equation}
  S\left(r,\theta,B_{\Phi}\right)=\alpha_0\,\bar{B}_{\Phi, {\rm bc}}(\theta)\,
  f_{\alpha}(r)\,g_{\alpha}(\theta)\;,
\end{equation}
with
\begin{eqnarray}
  f_{\alpha}(r)&=&\mbox{max}\,\left[0,\,1-\frac{(r-r_{\rm max})^2}
    {d_{\alpha}^2}\right]\\
  g_{\alpha}(\theta)&=&\frac{(\sin\theta)^2\cos\theta}
  {\mbox{max}\,\left[(\sin\theta)^2\cos\theta\right]}\\
  \bar{B}_{\Phi, {\rm bc}}(\theta)&=&\int_{r_{\rm min}}^{r_{\rm max}} 
      {\rm d}r\,h(r)\,B_{\Phi}(r,\theta)\label{B_bc}\;.
\end{eqnarray}
We use here $d_{\alpha}=0.05\,R_{\odot}$, which confines the poloidal source 
term above $r=0.935\,R_{\odot}$, peaking at $r_{\rm max}$. The function 
$h(r)$ is an averaging kernel
with $\int_{r_{\rm min}}^{r_{\rm max}} {\rm d}r\,h(r)=1$. We use a parabolic 
profile vanishing at $0.71\,R_{\odot}$ and $0.76\,R_{\odot}$ with peak at 
$0.735\,R_{\odot}$.

The boundary condition is $A=B_{\Phi}=0$ at the pole and
$\partial A/\partial\theta=B_{\Phi}=0$ at the equator, which selects the
dipole symmetry for the solution.
$B_{\Phi}$ vanishes at both radial boundaries. $A$ vanishes at the inner 
boundary and the poloidal field is assumed to be radial at the top boundary.
The dynamo part of our code has been compared intensively with the code
of \citet{Dikpati:Charbonneau:1999}.

We emphasize that the main goal of this paper is a fundamental 
understanding of dynamical effects caused by the feedback of the dynamo
generated field on differential rotation and meridional flow, rather than
a detailed model of the solar dynamo. We have chosen the dynamo
parameters and the parameters of differential rotation model such that the
solutions show a reasonable, but not detailed agreement in terms of butterfly 
diagram and dynamo period as well as amplitude of differential rotation and 
meridional flow with observations. We restrict our simulations to one
hemisphere and impose the dipole symmetry through our equatorial boundary
condition. We do not try to address the parity issue in this model, which has
been done for flux-transport dynamos by \citet{Dikpati:Gilman:2001:dynamo}.    

Before we discuss dynamic solutions, we present here as reference a kinematic
dynamo solution computed with the differential rotation and meridional flow
shown in Fig. \ref{f1}. We use for the $\alpha$-effect and amplitude of
$\alpha_0=0.125\,\vunit$ and include $\alpha$-quenching
with a quenching field strength of $1\,\mbox{T}$ ($10\,\mbox{kG}$). 
The butterfly diagram computed from the averaged toroidal field 
($\bar{B}_{\Phi, {\rm bc}})$
and the radial field at the 'surface' $r=0.985\,R_{\odot}$ is shown in Fig. 
\ref{f3}. The dynamo period for this setup is around $19$ years, the maximum
toroidal field at $r=0.735\,R_{\odot}$ is $1.28\,\mbox{T}$, the maximum field 
strength of the radial field at $r=0.985\,R_{\odot}$ is $0.01\,\mbox{T}$.

The butterfly diagram shows equatorward propagating activity belts starting 
around $50\degr$ latitude and having their peak field strength at around 
$40\degr$ latitude. The polar reversal of the poloidal field takes place 
during maximum activity with the toroidal field in low latitudes, changing 
sign from negative to positive while the low latitude toroidal field is 
positive as found in observations.

\section{Dynamo model with Lorentz force feedback}
\label{dynamicdynamo}
\subsection{General solution properties}

\begin{figure}
  \resizebox{\hsize}{!}{\includegraphics{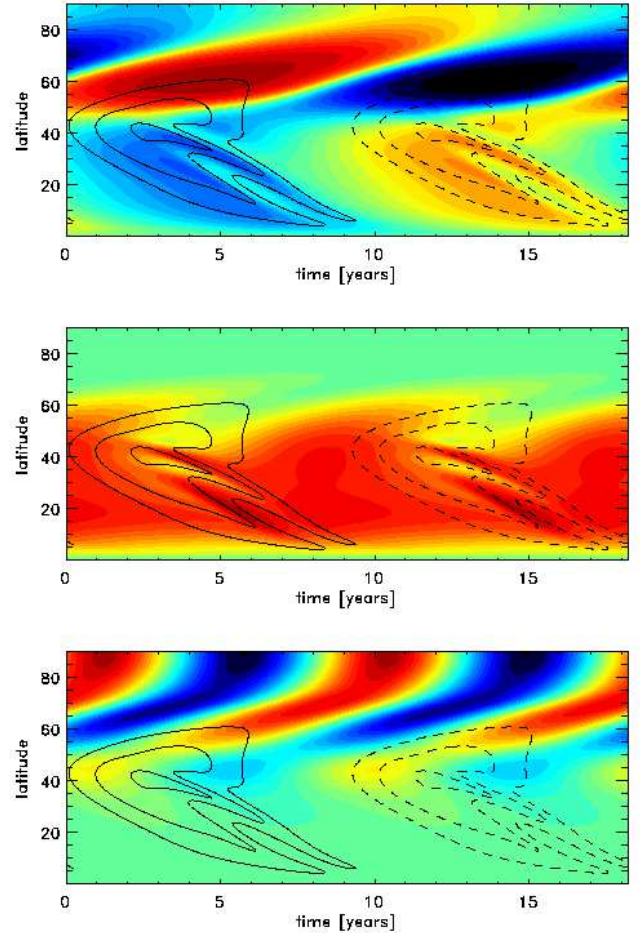}}
  \caption{Dynamo solution with Lorentz force feedback and no 
    $\alpha$-quenching. All dynamo parameters are the same as for the
    reference solution (with $\alpha$-quenching) shown in Fig. (\ref{f3}).
    Top panel: Butterfly diagram of toroidal magnetic field
    (contour lines) and radial field close to surface $r=0.985\,R_{\odot}$ 
    (color shades). The maximum toroidal field strength 
    is $1.17\,\mbox{T}$, the maximum radial field strength at the surface
    is $0.014\,\mbox{T}$. Middle panel: Meridional
    flow ($v_{\theta}$, color shades) at $r=0.75\,R_{\odot}$. The maximum 
    flow velocity is $2.27\,\vunit$. Bottom Panel: Torsional
    oscillation pattern ($\Omega-\bar{\Omega}$) at $r=0.985\,R_{\odot}$. 
    The maximum amplitude is $1\%$ of the core rotation rate corresponding
    to a variation of about $4\,\mbox{nHz}$. The torsional oscillation pattern
    at $r=0.735\,R_{\odot}$ is very similar, but has a slightly reduced 
    amplitude.
  }
  \label{f4}
\end{figure}

\begin{figure}
  \resizebox{8.5cm}{!}{\includegraphics{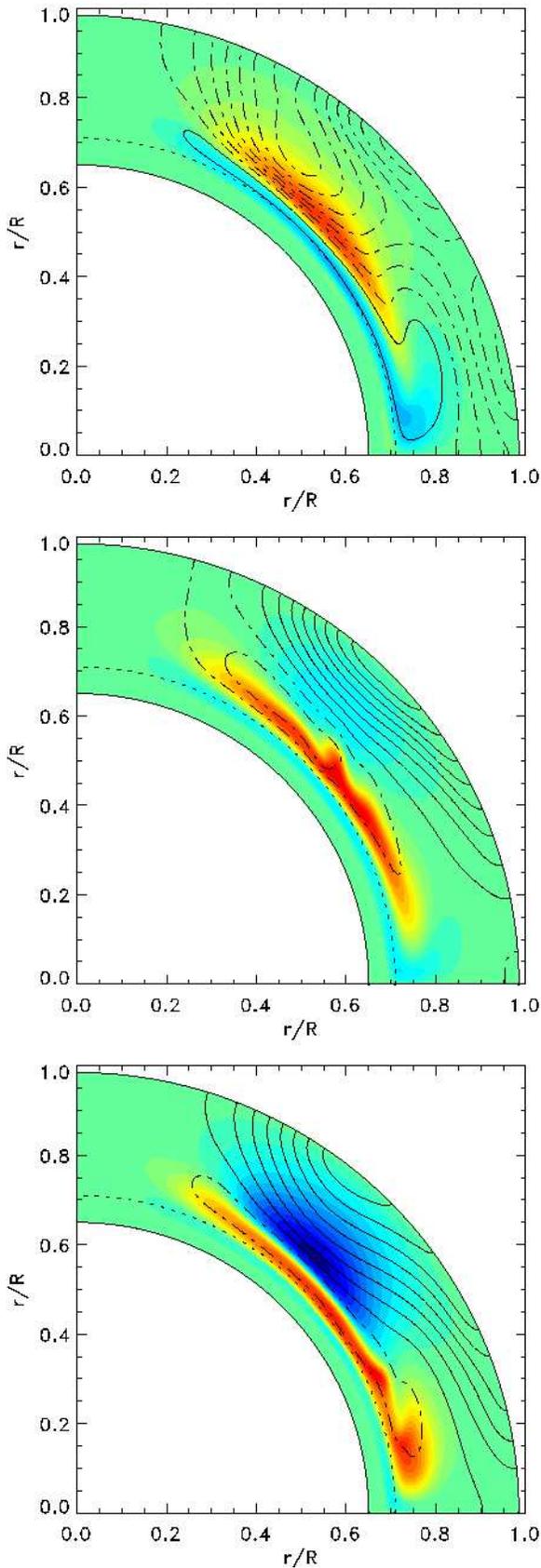}}
  \caption{Time evolution of magnetic field (color shades: toroidal field,
    contour lines: poloidal field lines). The frames shown correspond to 
    $t=0$, $t=3.25$, and $t=6.5$ in Figure \ref{f4}. The bottom panels
    shows clearly the breakup of the magnetic layer caused by magnetic 
    buoyancy.
  }
  \label{f5}
\end{figure}

\begin{figure}
  \resizebox{\hsize}{!}{\includegraphics{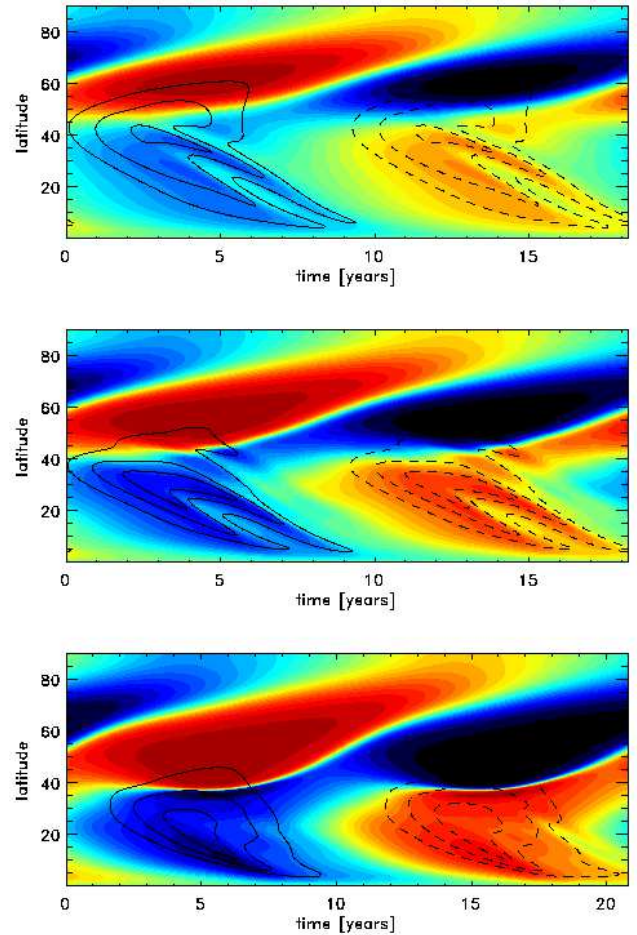}}
  \caption{Butterfly diagram (contour lines) and 
    radial field at $r=0.985\,R_{\odot}$ (color shades) for the solutions
    with $\alpha_0=$ $0.125\,\vunit$ (top), $0.25\,\vunit$ (middle), and 
    $0.5\,\vunit$ (bottom). The increase in Lorentz force 
    feedback leads to a confinement of magnetic activity to lower latitudes.
  }
  \label{f6}
\end{figure}

In this section we discuss results obtained with full Lorentz force
feedback on differential rotation and meridional flow. Since 
Lorentz force feedback introduces enough non-linearity to saturate
the dynamo, it is not necessary to include $\alpha$-quenching as typically
done in kinematic models. We present here three models varying the value
of the $\alpha$-effect in order to show different regimes in terms of
intensity of the Lorentz force feedback. The $\alpha_0$ values used are
$0.125$, $0.25$, and $0.5$ $\vunit$. We also present results
computed from a model with a magnetic diffusivity reduced in the bulk of the
convection zone from $\eta_{\rm cz}=10^8\,\dunit$ to 
$\eta_{\rm cz}=5\times 10^7\,\dunit$.

Figure \ref{f4} shows the model with $\alpha_0=0.125\,\vunit$.
The top panel displays the radial magnetic field close to the surface 
($r=0.985\,R_{\odot}$), the middle panel the meridional flow at 
$r=0.735\,R_{\odot}$, and the bottom panel the torsional oscillations at 
$r=0.985\,R_{\odot}$. In all three panels the toroidal field contours at 
$r=0.735\,R_{\odot}$ are indicated (solid: positive values, dashed: negative
values). Compared to Figure \ref{f3} the butterfly diagram shows major 
distortions caused by magnetic buoyancy breaking up the layer of toroidal 
field at the base of the convection zone. We present a more detailed 
discussion of the role of magnetic buoyancy in subsection \ref{magbuo}. 
The radial surface field is close
to that in Figure \ref{f3}. The equatorward meridional flow at 
$r=0.735\,R_{\odot}$ shows a variation of around $30\%$ of the mean flow
amplitude (around $2\,\vunit$) in anti-phase with the
toroidal field intensity, caused by the influence of the magnetic tension 
of the toroidal field. This feedback is not strong enough to switch off
the equatorward transport of magnetic field and therefore does not influence
the flux transport dynamo significantly. This is in agreement with 
\citet{Rempel:2005:transport}, who found that equatorward transport of
toroidal field is possible up to around $3\,\mbox{T}$. The bottom panel of
Figure \ref{f4} shows the temporal variation of the differential rotation
caused by the Lorentz force feedback, also known as torsional oscillations.
We find in our model mainly a poleward propagating oscillation pattern, 
starting at mid latitudes. The amplitude close to the surface is around $1\%$ 
of the core rotation rate, which corresponds to roughly $4$ nHz. We discuss
torsional oscillations in detail in section \ref{torsional}.

Fig. \ref{f5} shows 3 snapshots of the magnetic field (field lines of poloidal
field and toroidal field strength as color shades), corresponding to $t=0$,
$t=3.25$ and $t=6.5$ years in Fig. \ref{f4}. The middle and bottom panel
show the breakup of the magnetic layer caused by magnetic buoyancy force. 

Fig. \ref{f6} shows the dynamo solution (butterfly diagram and radial field 
close to surface) for the cases with increasing value of $\alpha_0$. In a 
kinematic model with $\alpha$-quenching as non-linearity, an increase of
$\alpha$ increases the field strength, but does not influence the shape of
the solution in great detail (with increased quenching the average profile
of $\alpha$ changes, which has a slight influence of the solution). In our
model the toroidal field shows only minor changes ($1.2$, $1.4$, and $1.2$
T for the cases with $\alpha_0$ values of $0.125$, $0.25$, and $0.5$ 
$\mbox{m}\,\mbox{s}^{-1}$, respectively), however the profile of the solution
changes significantly through the changes in meridional flow and differential
rotation. On average stronger feedback leads to a concentration of magnetic
activity to lower latitudes, which is the consequence of a strong cycle
variability of the meridional flow. While the average flow is roughly 
the same in all cases, the meridional flow tends to be more concentrated
toward the equator during the phase of the cycle when the poloidal field is 
transported downward, leading to the production of toroidal field at lower
latitudes, too. Even though interesting, this feature has most likely no
relevance to solar dynamos, since the degree of feedback required is not 
observed (e.g. torsional oscillations with an amplitude of around $20$ nHz).
When considering solutions with such a large degree of feedback magnetic 
buoyancy also plays an important role (subsection \ref{magbuo}).

\subsection{Energy flows within model}
\label{energyflow}

\begin{figure}
  \resizebox{\hsize}{!}{\includegraphics{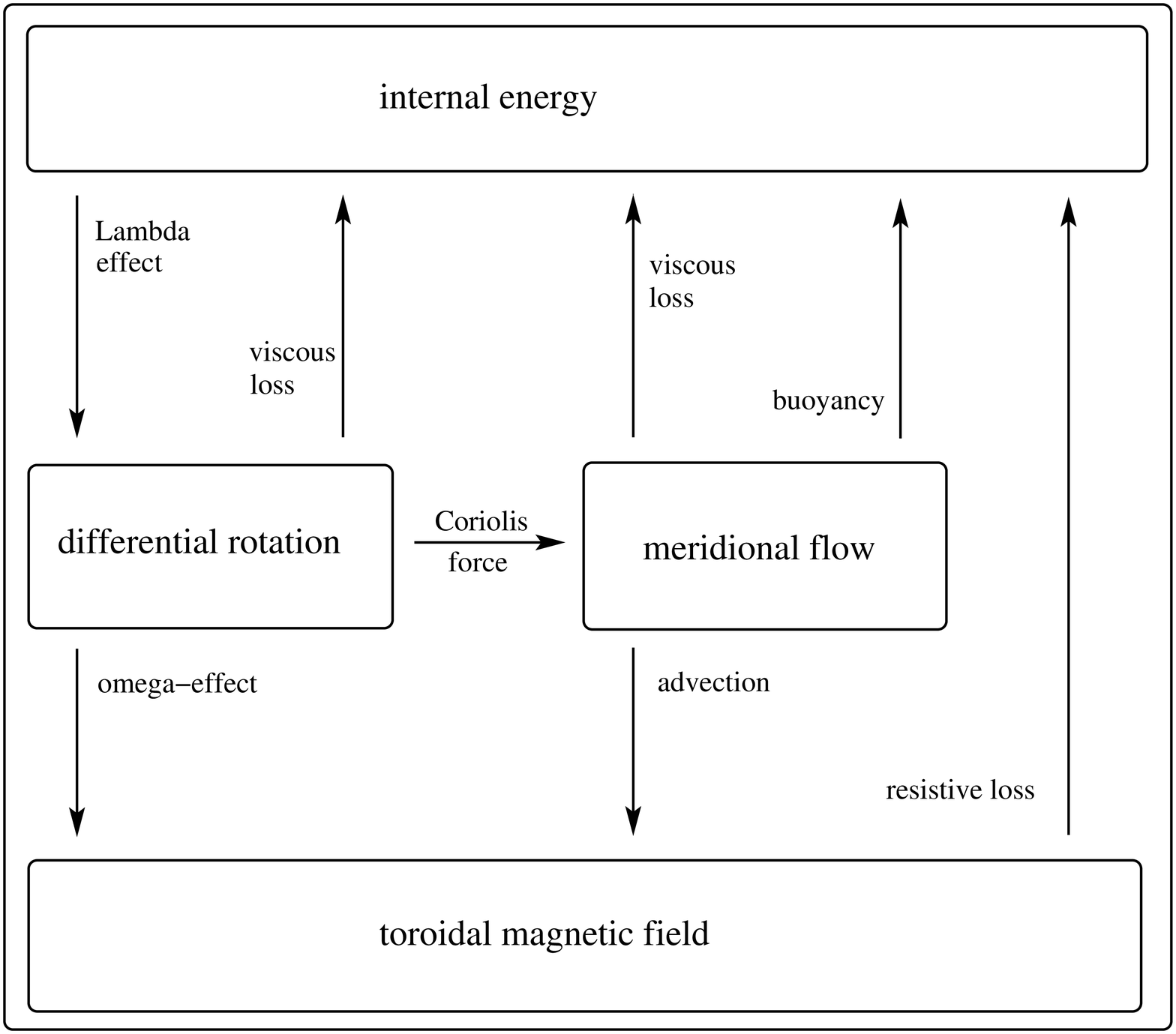}}
  \caption{Schematic view of energy fluxes in the coupled 
    dynamo-differential rotation model.
  }
  \label{f7}
\end{figure}

\begin{deluxetable*}{cccccccccc}
  \tablecaption{Summary of results\label{tab1}}
  \tablehead{ \colhead{quantity} & \colhead{unit} & \colhead{reference} &
    \multicolumn{2}{c}{$\alpha_0=0.125\,\vunit$} &
    \multicolumn{2}{c}{$\alpha_0=0.25\,\vunit$} &
    \multicolumn{2}{c}{$\alpha_0=0.5\,\vunit$} & 
    \colhead{$\eta_{\rm cz}=5\times 10^7\,\dunit$}
  }
  \startdata
  $\left(\bar{\Omega}_{\rm eq}-\bar{\Omega}_{\rm pole}\right)/\Omega_0$ &
  & $0.27$ & $0.21$ & $(0.21)$ & $0.15$ & $(0.15)$ & $0.1$ & $(0.12)$ &$0.15$  
  \vspace{0.1cm}\\
  $\mbox{max}(\Omega-\bar{\Omega})_{90\degr}$ & [nHz]
  & $\ldots$ & $4.7$ & $(5.8)$ & $11.5$ & $(11.1)$ &$17.7$ &$(24.2)$ &$12.9$   
  \vspace{0.1cm} \\
  $\mbox{max}(\Omega-\bar{\Omega})_{60\degr}$ & [nHz]
  & $\ldots$ & $3.5$ & $(2.9)$ & $6.9$ & $(5.8)$ & $8.3$ & $(12.1)$ & $9.7$   
  \vspace{0.1cm} \\
  $\mbox{max}(B_{\Phi})$ & $[\mbox{T}]$        
  & $\ldots$ & $1.2$ & $(1.2)$ & $1.4$ & $(1.1)$ & $1.2$ & $(1.2)$ & $1.5$  
  \vspace{0.1cm}\\
  $\mbox{max}(B_{r})$ & $[\mbox{T}]$          
  & $\ldots$ & $0.014$ & $(0.015)$ & $0.022$ & $(0.021)$ & $0.027$ & 
  $(0.047)$ & $0.02$ 
  \vspace{0.1cm}\\
  $\bar{E}_B$ & $[10^{31}\mbox{J}]$
  & $\ldots$ & $2.8$ & $(3.5)$ & $4.6$ & $(4.4)$ & $5.1$ & $(4.6)$ & $4.9$  
  \vspace{0.1cm} \\
  $\mbox{max}\left[\left(E_B-\bar{E}_B\right)/\bar{E}_B\right]$  &
  & $\ldots$ & $0.12$ & $(0.15)$ &  $0.22$ & $(0.26)$ & $0.26$ & $(0.55)$ & 
  $0.21$
  \vspace{0.1cm} \\   
  $\mbox{max}(E_B^{+})_{\rm bc}$ & $[10^{31}\mbox{J}]$
  & $\ldots$ & $1.7$ & $(2.1)$ & $2.2$ & $(2.2)$ & $2.1$ & $(2.7)$ & $2.4$  
  \vspace{0.1cm} \\
  $\mbox{max}(\Phi^{+})_{\rm bc}$ & $[10^{16}\mbox{Wb}]$
  & $\ldots$ & $1.1$ & $(1.3)$ & $1.2$ & $(1.3)$ & $1.2$ & $(1.6)$ & $1.2$  
  \vspace{0.25cm} \\
  \colrule\\
  $Q_{\Lambda}$ & $[F_{\odot}]$ 
  & $0.014$ & $0.013$ & $(0.013)$ & $0.011$ & $(0.01)$ & $0.008$ & 
  $(0.008)$ & $0.012$   
  \vspace{0.1cm} \\
  $Q_{\nu}^{\Omega}$ & $[Q_{\Lambda}]$
  & $0.574$ & $0.5$ & $(0.484)$ & $0.405$ & $(0.415)$ & $0.304$ & 
  $(0.357)$ & $0.459$ 
  \vspace{0.1cm} \\
  $Q_{C}$ &  $[Q_{\Lambda}]$         
  & $0.425$ & $0.43$ & $(0.429)$ & $0.445$ & $(0.423)$ & $0.465$ & 
  $(0.412)$ & $0.441$  
  \vspace{0.1cm}  \\
  $Q^{\Omega}_L$ &  $[Q_{\Lambda}]$ 
  & $\ldots$ & $0.069$ & $(0.087)$ & $0.149$ & $(0.162)$ & $0.232$ & 
  $(0.231)$ & $0.1$  
  \vspace{0.1cm}  \\
  $Q_B$ & $[Q_{\Lambda}]$      
  & $0.419$ & $0.41$ & $(0.404)$ & $0.414$ & $(0.384)$ & $0.423$ & 
  $(0.357)$ & $0.409$  
  \vspace{0.1cm} \\
  $Q^M_{\nu}$ &    $[Q_{\Lambda}]$         
  & $0.005$ & $0.005$ & $(0.005)$ & $0.007$ & $(0.007)$ & $0.01$ & 
  $(0.01)$ & $0.008$   
  \vspace{0.1cm} \\
  $Q^M_L$ & $[Q_{\Lambda}]$   
  & $\ldots$ & $0.014$ & $(0.019)$ & $0.024$ & $(0.032)$ & $0.033$ & 
  $(0.043)$ & $0.023$   
  \vspace{0.1cm} \\
  $Q_{\eta}$ &   $[Q_{\Lambda}]$     
  & $\ldots$ & $0.083$ & $(0.104)$ & $0.172$ & $(0.192)$ & $0.268$ & 
  $(0.272)$ & $0.123$  
  \vspace{0.1cm} \\
  $Q_{\eta}$ &   $[F_{\odot}]$     
  & $\ldots$ & $0.0011$ & $(0.0014)$ & $0.0019$ & $(0.0019)$ & 
  $0.0021$ & $(0.0022)$ & $0.0015$  
  \vspace{0.1cm} 
  \enddata
  \tablecomments{Summary of dynamo simulations discussed. The top portion
  shows a few global properties of the solution, the bottom portion the
  energy transfer terms defined by Eqs. (\ref{qlambda}) - (\ref{qeta}).
  Values in brackets refer to models in which the magnetic buoyancy
  term is switched off (see subsection \ref{magbuo} for further details). 
  The maximum of $\Omega-\bar{\Omega}$ and $B_r$ is evaluated at 
  $0.985\,R_{\odot}$, the maximum of $B_{\Phi}$ at $0.735\,R_{\odot}$.
  $\mbox{max}(E_B^{+})_{\rm bc}$ and $\mbox{max}(\Phi^{+})_{\rm bc}$ are 
  the maximum values of magnetic energy and magnetic flux integrated over the
  interval $[0.71, 0.76]\,R_{\odot}$, considering only one toroidal field 
  polarity. The value of $Q_{\Lambda}$ and $Q_{\eta}$ (last row) are given
  relative to the solar luminosity, all the other energy exchange terms
  are relative to $Q_{\Lambda}$. Note
  that the accuracy of the energy exchange terms is around $0.001$, so the
  equilibrium relations $Q_{\Lambda}=Q^{\Omega}_{\nu}+Q_{C}+Q^{\Omega}_{L}$,
  $Q_{C}=Q^M_{\nu}+Q_{B}+Q^{M}_{L}$, and $Q_{\eta}=Q^{\Omega}_{L}+Q^{M}_{L}$
  are only fulfilled within that error margin.}
\end{deluxetable*}

In this section we discuss the dynamo solutions by analyzing the
energy flows between the different energy reservoirs of the model. This
allows us to understand the saturation mechanism of the non-linear dynamo 
on a more quantitative level.

Figure \ref{f7} shows a schematic of the energy flows in our coupled
dynamo-differential rotation model. For reasons of simplicity we consider 
here only the energy of the toroidal field, since the energy of the poloidal 
field plays only a minor role. The following equations describe the change
of rotation energy $E_{\Omega}$, energy of meridional flow $E_M$, and
energy of toroidal magnetic field $E_B$:
\begin{eqnarray}
  \frac{\partial E_{\Omega}}{\partial t}&=&Q_{\Lambda}-Q^{\Omega}_{\nu}
  -Q_{C}-Q^{\Omega}_{L}\label{erot}\\
 \frac{\partial E_M}{\partial t}&=&Q_{C}-Q^M_{\nu}-Q_{B}
  -Q^{M}_{L}\label{emerid}\\
  \frac{\partial E_B}{\partial t}&=&Q^{\Omega}_{L}+Q^{M}_{L}
  -Q_{\eta}\label{emag}\;,
\end{eqnarray}
where the energy reservoirs are given by
\begin{eqnarray}
  E_{\Omega}&=&\int\mbox{d}V\,\frac{1}{2}\rh_0 s^2 \Omega^2\\
  E_M&=&\int\mbox{d}V\,\rh_0\frac{\vec{v}_m^2}{2}\\
  E_B&=&\int\mbox{d}V\,\frac{B_{\Phi}^2}{2\,\mu_0}
\end{eqnarray}
and the exchange terms are given by
\begin{eqnarray}
  Q_{\Lambda}&=&-\int\mbox{d}V\,\nu_t\rh_0 s
     \left(\frac{\partial\Omega}{\partial r}
     \Lambda_{r\phi}+\frac{1}{r}\frac{\partial\Omega}{\partial \theta}
     \Lambda_{\theta\phi}\right)\label{qlambda}\\ 
  Q^{\Omega}_{\nu}&=&\int\mbox{d}V\,\nu_t\rh_0 s^2 \left[
     \left(\frac{\partial\Omega}{\partial r}\right)^2+
     \left(\frac{1}{r}\frac{\partial\Omega}{\partial \theta}\right)^2\right]\\
  Q_{C}&=&-\int\mbox{d}V\,s^2\rh_0\vec{v}_m\cdot\vec{\nabla}
     \frac{\Omega^2}{2}\\ 
  Q^{\Omega}_{L}&=&-\int\mbox{d}V\,\Omega\,\vec{B}_p\cdot\vec{\nabla}
     \left(s B_{\Phi}\right)\\ 
  Q^M_{\nu}&=&\int\mbox{d}V\,\frac{1}{2}\left[R_{rr}{E}_{rr}+2{R}_{r\theta}
     {E}_{r\theta}+{R}_{\theta\theta}{E}_{\theta\theta}\right.\nonumber\\
     &&\left.+{R}_{\phi\phi}{E}_{\phi\phi}\right]\\
  Q_{B}&=&-\int\mbox{d}V\,v_r\rh_0 g\frac{s_1}{\gamma}\label{qentr}\\
  Q^{M}_{L}&=&\int\mbox{d}V\,\frac{B_{\Phi}}{s}\,\vec{v}_m\cdot\vec{\nabla}
     \left(s B_{\Phi}\right)\\  
  Q_{\eta}&=&\int\mbox{d}V\,\frac{\eta_t}{s^2}\left[\left(\frac{
     \partial \left(s B_{\Phi}\right)}{\partial r}\right)^2
     +\left(\frac{1}{r}\frac{\partial \left(s B_{\Phi}\right)}
     {\partial \theta}\right)^2\right]\label{qeta}  
\end{eqnarray}
Here $s=r\sin\theta$ denotes the distance to the axis of rotation and
\begin{equation}
  \int\mbox{d}V=4\pi\int_{r_{\rm min}}^{r_{\rm max}}\mbox{d}r\int_{0}^{\pi/2}
  \mbox{d}\theta r^2\sin\theta
\end{equation}
denotes the integral over the entire volume of the sphere from $r=r_{\rm min}$
to $r=r_{\rm max}$. We emphasize that we solve our model only for the
northern hemisphere but we compute from that the energy conversion for
the entire sphere. The quantity $\vec{v}_m=(v_r,v_{\theta},0)$ denotes 
the meridional flow and $\vec{B}_p=(B_r,B_{\theta},0)$ the poloidal magnetic
field.

\citet{Rempel:2005:random} showed already the derivation for the non-magnetic
part of the system Eqs. (\ref{erot}) - (\ref{emag}). The exchange terms 
Eqs. (\ref{qlambda}) - (\ref{qeta}) are not unique 
expressions since they can be transformed in the form:
\begin{equation}
  expression_1=\vec{\nabla}\cdot\vec{flux}+expression_2\;,
\end{equation}
provided that the volume integral over the flux divergence vanishes. We use 
the appropriate closed boundary conditions for all variables except for 
$B_{\Phi}$, however, the resistive flux across the upper boundary turns out
to be negligible. We also emphasize that
we defined a few of the exchange terms with an opposite sign than in
\citet{Rempel:2005:random}. Here all $Q_{\ldots}^{\ldots}$ are positive
so that the signs in Eqs. (\ref{erot}) - (\ref{emag}) clearly state which terms
are sources and which are sinks for the corresponding energy reservoir.

The third column of table \ref{tab1} summarizes the energy flow of the 
differential rotation 
reference model we use for all dynamo simulations. In this model the
$\Lambda$-effect converts $1.4\%$ of the solar energy flux into rotation
energy. $57.4\%$ of this energy is lost directly through viscous dissipation
of the differential rotation,
while $42.5\%$ is flowing into the reservoir of meridional flow by means
of the Coriolis force. The major fraction of this amount returns into
the reservoir of internal energy through work against buoyancy force, only
a small fraction $<1\%$ through viscous dissipation of the meridional flow.
In the reference model the pole-equator difference in rotation rate is 
$27\%$ of the core rotation rate. 

Columns 4 to 10 of table \ref{tab1} summarize the results obtained from 
the dynamo models with different $\alpha$ and $\eta_{\rm cz}$ values. The 
top portion of Table \ref{tab1} shows the pole-equator difference in rotation 
rate, the amplitude of torsional oscillations, the maximum toroidal and radial 
field strength, average magnetic energy and fluctuation of magnetic energy
over a dynamo cycle as well as maximum magnetic energy and flux of one 
polarity at the base of the convection zone. The bottom portion shows the 
energy exchange terms Eqs. (\ref{qlambda}) to (\ref{qeta}) averaged over a 
cycle ($12$ cycles in case of the irregular solutions shown in Fig. \ref{f8}). 
The quantity $Q_{\Lambda}$ is given 
in units of the solar energy flux and the following terms relative to 
$Q_{\Lambda}$. The last line gives $Q_{\eta}$ in units of the solar energy 
flux. 
 
As already mentioned in the previous section, increasing the value of
$\alpha$ does not lead to a significant increase of the toroidal field 
strength. The same applies to the magnetic energy and the magnetic flux at the
base of the convection zone. At the same time the equator-pole difference of 
the differential rotation decreases monotonically from $0.21\Omega_0$ in the
case with $\alpha_0=0.125\,\vunit$ to $0.1\Omega_0$ in the case with
$\alpha_0=0.5\,\vunit$. For comparison, the reference model has an
equator-pole difference of $0.27\Omega_0$. This shows clearly that the 
saturation mechanism of the dynamo is the reduction of the shear through
Lorentz force feedback on differential rotation. This becomes also evident by
looking at the energy exchange terms $Q^{\Omega}_{\nu}$ and $Q^{\Omega}_L$:
While $Q^{\Omega}_{\nu}$ is decreasing and $Q^{\Omega}_L$ increasing, the sum
of both relative to $Q_{\Lambda}$ remains roughly the same, meaning that in
the dynamo solutions viscous stress is replaced by Maxwell stress. The energy
transfer to meridional flow $Q_C$ shows only a weak increase with $\alpha_0$.
The energy transferred in total into magnetic energy (and dissipated back to
internal energy), $Q_{\eta}$ increases relative to $Q_{\Lambda}$ from
$8.3\%$ to $26.1\%$ with increasing $\alpha_0$. Even though we use a fixed 
parameterization for the $\Lambda$-effect in our model, the energy which is 
converted by the $\Lambda$-effect ($Q_{\Lambda}$) decreases with increasing 
$\alpha_0$, since in Eq. (\ref{qlambda}) also the shear enters. As a 
consequence, $Q_{\eta}$ does not change that much in absolute units
(last row in Table \ref{tab1}) when $\alpha_0$ is increased from 
$0.25\,\vunit$ to $0.5\,\vunit$. This is consistent with the almost constant 
value of the magnetic energy in both cases. 

A different way to look at the saturation is as follows. In the model with
$\alpha_0=0.5\,\vunit$ the radial surface field (a measure for the poloidal
field strength) almost doubled compared to the model with 
$\alpha_0=0.125\,\vunit$. At the same time equator-pole difference of $\Omega$
is reduced to roughly half the value, leading to the same toroidal field 
strength through the $\Omega$-effect. Since the Lorentz force is proportional
to the product of poloidal and toroidal field, the feedback on $\Omega$
has almost doubled. This leaves however the question if it is possible to
obtain larger toroidal field strength by going in the opposite direction
and reducing the poloidal field strength. Inspecting Eq. (\ref{bphi})  gives
and estimate of the toroidal/poloidal field ratio, which can be obtained 
through the $\Omega$-effect:
\begin{equation}
  \frac{B_{\Phi}}{B_{\theta}}\sim \tau \frac{\partial \Omega}{\partial \theta}
    \sim 100\;,\label{field-ratio}
\end{equation} 
where we used $\tau=5\,\mbox{years}$ and $\partial \Omega/\partial \theta
\sim 0.2\,\Omega_0$. For the model with $\alpha_0=0.125\,\vunit$ 
$B_{\theta}$ within the convection zone peaks at around $0.03\,\mbox{T}$,
which does not allow for much larger toroidal field keeping in mind the 
reduction of the poloidal field in this thought experiment and considering that
the estimate Eq. (\ref{field-ratio}) does not include resistive loss during
the amplification phase.

The cycle variation of $\Omega$ (torsional oscillation) increases roughly 
proportional to $\alpha_0$ from around $4.7$ nHz to $18$ nHz at the pole,
the change at $60\degr$ latitude is about half that value. Given the 
observational constraint that solar torsional oscillations at $60\degr$
have an amplitude of around $2$ nHz, the solar dynamo is most likely 
operating in a regime with weak Lorentz force feedback, close to our model 
with $\alpha_0=0.125\,\vunit$ or even smaller.
In that model the amount of energy converted to
magnetic energy is around $8\%$ of the energy converted into differential
rotation by the $\Lambda$-effect, in absolute units, around $0.1\%$ of the
solar energy flux. This number is interestingly very close to the variation
of solar irradiance throughout the solar cycle.

In the last column of Table \ref{tab1} we show results obtained from a 
model with $\alpha_0=0.125\,\vunit$, but magnetic diffusivity of only
$5\times 10^7\,\dunit$ in the convection zone. This model is very close to
the model with $\alpha_0=0.25\,\vunit$ and a diffusivity of $10^8\,\dunit$.
The magnetic field strength as well as magnetic energy is slightly higher,
at the same time the amount of energy extracted from differential rotation
is lower due to the lower magnetic diffusivity. Nevertheless, the amplitude
of torsional oscillations is larger compared to the model with higher 
diffusivity.

\subsection{Saturation field strength of dynamo}
Various models of rising magnetic flux-tubes \citep{Choudhuri:Gilman:1987,
Fan:etal:1993,Schuessler:etal:1994,Caligari:etal:1995,Caligari:etal:1998}
inferred a field strength of around $10\,\mbox{T}$ at the base of the
convection zone, which is almost one magnitude more than the field strength
we obtain in our model. Since our model is a mean field model, the mean field
strength is not necessarily identical to the field strength of individual flux
elements in case of an intermittent field. When we compare our mean field 
solution to a solution with an intermittent field it is required to compare
energy densities (which also ensure the same Lorentz force densities) rather
than average field strength, since this ensures the same level of dynamic 
feedback. Suppose an intermittent field with a filling 
factor $f$ and individual flux elements with field strength $B_f$. Conserving
the mean energy leads a relation between mean field strength $B$ and $B_f$ 
of $B_f\sim B/\sqrt{f}$. For
a filling factor of $f=0.1$ this would allow for around $4\,\mbox{T}$ for
individual flux elements. Another way to estimate maximum field strength
is using the field energy directly. If we assume that the toroidal magnetic 
field is stored at the base of the convection zone close to the equator
in a layer with a thickness $\Delta r$ and a latitudinal extent $\Delta l$, 
the magnetic energy is 
$E_B\sim 2 \pi r \Delta r \Delta l B^2/(2\mu_0)=\pi r \Phi B/\mu_0$, with the
magnetic flux $\Phi$. Observations lead to an estimate of the
magnetic flux of $10^{16} - 10^{17}\,\mbox{Wb}=10^{24} - 10^{25}\,\mbox{Mx}$ 
\citep{Galloway:Weiss:1981,Schrijver:Harvey:1994}; however,
the magnetic flux produced during a cycle at the base of the convection zone 
is not necessarily identical to the observed surface flux. On the one hand 
a rising flux rope could produce more than one spot group (which would allow 
for less flux at the base of the convection zone) on the other hand it is 
also likely that only a fraction of the flux at the base of the convection 
zone shows up in spots at the surface. The flux value we mention in Table
\ref{tab1} is the maximum flux of the dominant polarity available at a given 
time, while the estimates from observations are the cumulative values of flux
integrated over the cycle.

If we use the average energy conversion rate of $0.001\,F_{\odot}$
and integrate that over a cycle length of $11\,\mbox{years}$ 
we end up with an upper (very optimistic) estimate of 
$1.3\times 10^{32}\,\mbox{J}$, not considering any dissipative loss. In this 
case a value of $\Phi=10^{16}\,\mbox{Wb}$ would allow for $B=10\,\mbox{T}$,
the higher value of $\Phi=10^{17}\,\mbox{Wb}$ for only $B=1\,\mbox{T}$.
Accounting for the  additional dissipative loss, a field strength of more 
than a few T seems unlikely unless the average energy conversion rate would 
be larger. In our model that would lead to a contradiction with the observed
amplitude of torsional oscillations.

It has been pointed out by \citet{Rempel:Schuessler:2001} that also potential
energy of the superadiabatic convection zone can be used for the amplification 
of magnetic field. In contrast to the amplification through the 
$\Omega$-effect (shearing up of poloidal field by differential rotation), 
this does not lead to a feedback on differential rotation. Recently
Y. Fan (2006, private communication) repeated simulations of rising flux 
tubes using an anelastic 3D MHD code. These simulations are therefore not 
bound to the thin flux tube approximation previously used. Preliminary results 
show that the low latitude emergence and the observed asymmetries between 
leading and following spots (the tilt angle is a more complicated problem
due to the influence of Coriolis force and twist of the flux tube, which is
currently being investigated) can be reproduced with magnetic flux tubes 
having an initial field strength only around $3\,\mbox{T}$, which would 
therefore relax the constraint on the field strength at the base of the 
convection zone.
    
\subsection{Role of magnetic buoyancy}

\label{magbuo}

\begin{figure}
  \resizebox{\hsize}{!}{\includegraphics{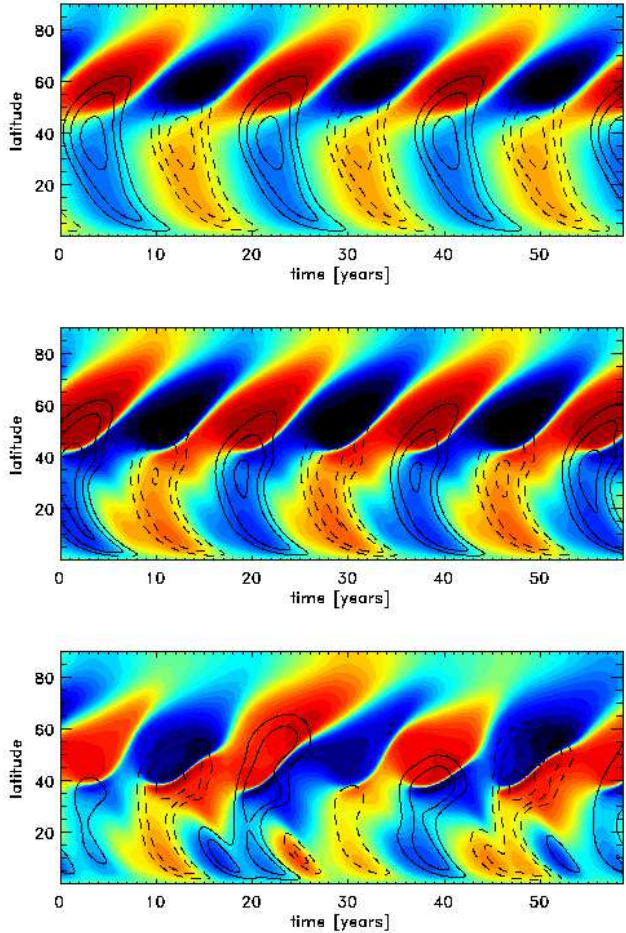}}
  \caption{Magnetic field evolution for models with magnetic buoyancy switched
    off. Similar to Fig. \ref{f6} solutions
    with $\alpha_0=$ $0.125\,\vunit$ (top), $0.25\,\vunit$ (middle), and 
    $0.5\,\vunit$ (bottom) are shown. 
  }
  \label{f8}
\end{figure}

Our model includes magnetic buoyancy in two different ways: implicit buoyancy
as part of the parametrized Babcock-Leighton $\alpha$-effect and explicit
(resolved) buoyancy resulting from solving the full momentum equation. While
the first effect is essential for the dynamo model, the latter leads mainly to 
a distortion of the toroidal magnetic field shown in Figs.~\ref{f4} to 
\ref{f6} and is not essential for the operation of the model. Having 
parametrized and resolved buoyancy together in a mean field model is of 
certain conceptual concern since it washes out the boundary between resolved
and parametrized processes. Another point of concern is the fact that most
buoyancy instabilities are non-axisymmetric and therefore our axisymmetric 
model does not capture the most unstable modes. One way of evaluating the 
importance of the explicit buoyancy in this model is to neglect the term 
$\sim p_{\rm mag}$ in Eq. (\ref{vrad}) that is responsible for it. This is
equivalent to adding to $s_1$ a perturbation of $-p_{\rm mag}/p_0$, which means
physically that the toroidal field is stored in a buoyancy free equilibrium 
at the base of the convection zone. The Babcock-Leighton $\alpha$-effect 
addresses buoyancy instabilities below the resolved scale and is therefore
not in contradiction with this assumption. Since this additional entropy
perturbation is not considered in Eq. (\ref{entr}) this leads to a small 
inconsistency in the entropy equation. For the analysis of the energy fluxes
we use instead of Eq. (\ref{qentr})
\begin{equation}
  Q_{B}=-\int\mbox{d}V\,v_r\rh_0 g\frac{1}{\gamma}
  \left(s_1-\frac{p_{\rm mag}}{p_0}\right)\;.
\end{equation}
Fig. \ref{f8} shows the magnetic field evolution similar to Fig. (\ref{f6}) 
with magnetic buoyancy switched off. While the top panel (solution with 
$\alpha_0=0.125\,\vunit$) is very close to the kinematic reference solution
with $\alpha$-quenching, the middle and bottom panels show dynamo solutions
with irregular cycles due to the non-linear feedback. Despite the significant
difference in the magnetic field pattern and temporal evolution, the 
differences in the energy exchange terms are not that significant. The most
obvious differences occur in the model with $\alpha_0=0.5\,\vunit$ in terms
of amplitude of torsional oscillations, radial surface field strength, and
variability of magnetic energy. The almost a factor of $2$ larger radial field
is also the reason for the large irregularity of the solution. The stronger
poloidal field at the surface leads through the Lorentz force to a stronger 
variation in the meridional flow that changes the latitude at which the 
magnetic field is transported downward. This in return changes the latitudinal 
extent and strength of the next cycle in a way that a periodic solution is 
not possible anymore. If buoyancy is considered the radial surface field does
not reach the threshold required for a highly irregular solution.

\section{Torsional oscillations}
\label{torsional}

\begin{figure}
  \resizebox{\hsize}{!}{\includegraphics{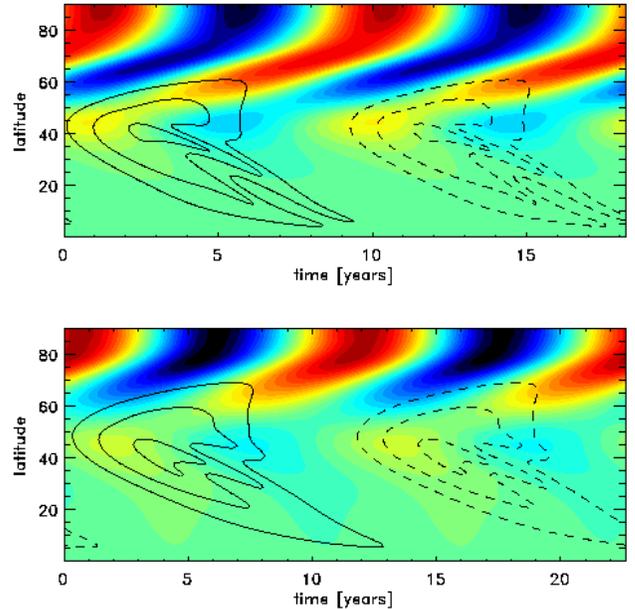}}
  \caption{Influence of meridional flow profile on phase relation between
    high latitude torsional oscillation and low latitude toroidal magnetic
    field. Top: solution shown in Fig. \ref{f4} (bottom panel) for reference;
    bottom: model with meridional flow returning in higher latitudes
    (n=2 instead of n=3 in reference model).
  }
  \label{f9}
\end{figure}

Solar torsional oscillations have been known to exist for more than two 
decades. \citet{Howard:Labonte:1980} presented the first observations of
torsional oscillations using Mt. Wilson Doppler measurements and
pointed out the 11 year periodicity and the relation to the solar
cycle. These early observations showed only the equatorward 
propagating branch at low latitudes. The high latitude branch 
(above $60\degr$), which is in amplitude at least twice as strong as the 
equatorward propagating branch, was found more recently through helioseismic 
measurements by \citet{Toomre:etal:2000,Howe:etal:2000:solphys,Antia:Basu:2001,
Vorontsov:etal:2002,Howe:etal:2005}. These inversions also show that the high
latitude signal penetrates almost all the way to the base of the convection 
zone. The depth penetration of the low latitude signal is more uncertain due
to the lower amplitude that is comparable to the uncertainties of the inversion
methods in the lower half of the convection zone. 

\subsection{Mechanical forcing of torsional oscillations}
The models discussed so far can only explain the polar branch of the
torsional oscillation pattern as a very robust result through the feedback of
mean field Lorentz force on differential rotation (mechanical forcing). 
We also showed that
some dynamo models lead to quite significant amplitudes of these oscillations,
contrary to observation, which therefore impose constraints on dynamo 
parameters (amplitude of $\alpha$-effect and the magnetic diffusivity in
convection zone).
Other information that can be extracted from torsional oscillations is 
their phase relative to the magnetic cycle. The phase relation shown in
Fig. 3 of \citet{Vorontsov:etal:2002} is such that the pole is rotating faster
during solar minimum and slower during maximum, when  solar activity peaks
around $15\degr$ latitude. In order to determine the phase relation in our
model, we have to define what phase corresponds to 'solar minimum' and
'solar maximum', given a butterfly diagram which is close to, but not exactly,
solar like. If we define 'solar maximum' through the maximum field strength
at the base of the convection zone, the location of the activity belt is at
around $40\degr$ and the torsional oscillation pattern changes from faster 
to slower rotation at the pole during that time. If we however define 
'solar maximum' as the time where the activity belt is around $15\degr$ 
latitude, the phase relation of the model matches the observed torsional 
oscillations.
Since the phase in a flux-transport dynamo is a consequence of the advection 
of field by the meridional flow, while the amplitude depends on details of
microphysics that might not be well represented or even missing in this 
mean field model (e.g. a tachocline $\alpha$-effect), the
latter definition is most likely more robust and should be used when 
comparing results to the sun. The phase relation is indeed tied to the 
structure of the meridional flow. Fig. \ref{f9} compares the model shown in 
Fig. \ref{f4} with model that has a meridional flow returning at higher
latitudes (we used instead of $n=3$ and $\Lambda_0=1$ $n=2$ and $\Lambda_0=0.8$
in the reference model). While in the original model slower rotation at the
pole coincided with an activity belt location at around $20\degr$, it coincides
with an activity belt location at around $30\degr$ in the latter model.
Also note that the dynamo period of the latter model is around $22.5$ years
instead of $18$ years, due to the longer overturning time of the meridional 
flow.  

\subsection{Thermal forcing of torsional oscillations}
An alternative explanation for the low latitude branch of the torsional 
oscillations was given by \citet{Spruit:2003}: Enhanced surface cooling in 
the active region belt leads to a pressure imbalance that drives a geostrophic 
flow at the edges of the active region belt showing the properties of the 
observed pattern . 
The low latitude oscillation pattern would be therefore purely surface driven
and not a signature of Lorentz force feedback within the convection zone.

\begin{figure}
  \resizebox{\hsize}{!}{\includegraphics{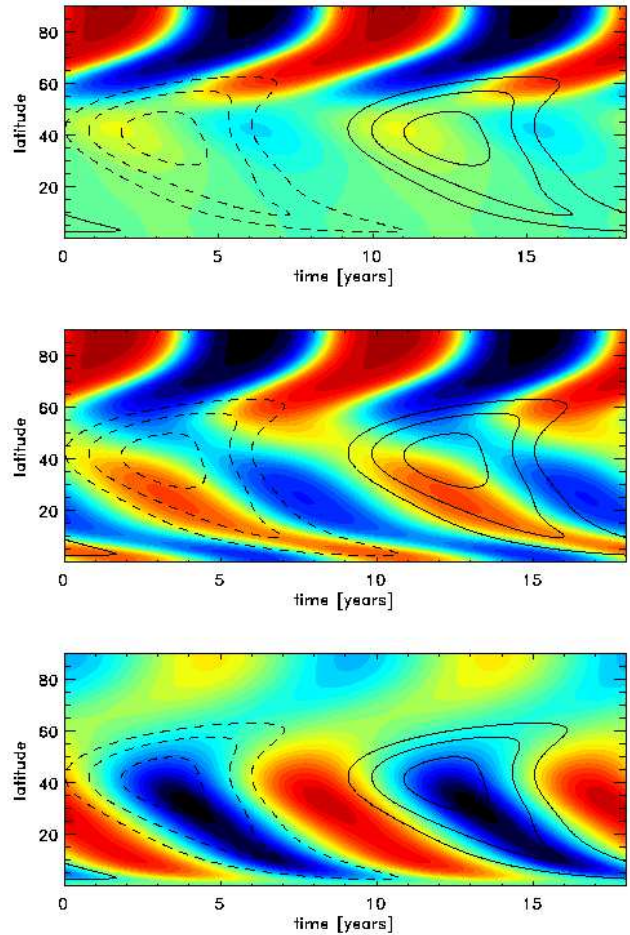}}
  \caption{Torsional oscillations caused through enhanced radiative losses in
    the active region belt. Top: reference model with no surface cooling;
    middle: model with surface cooling, the maximum amplitude is $4nHz$, 
    bottom: surface temperature variation, the maximum amplitude is $0.2$ K.
  }
  \label{f10}
\end{figure}
\begin{figure}
  \resizebox{8.5cm}{!}{\includegraphics{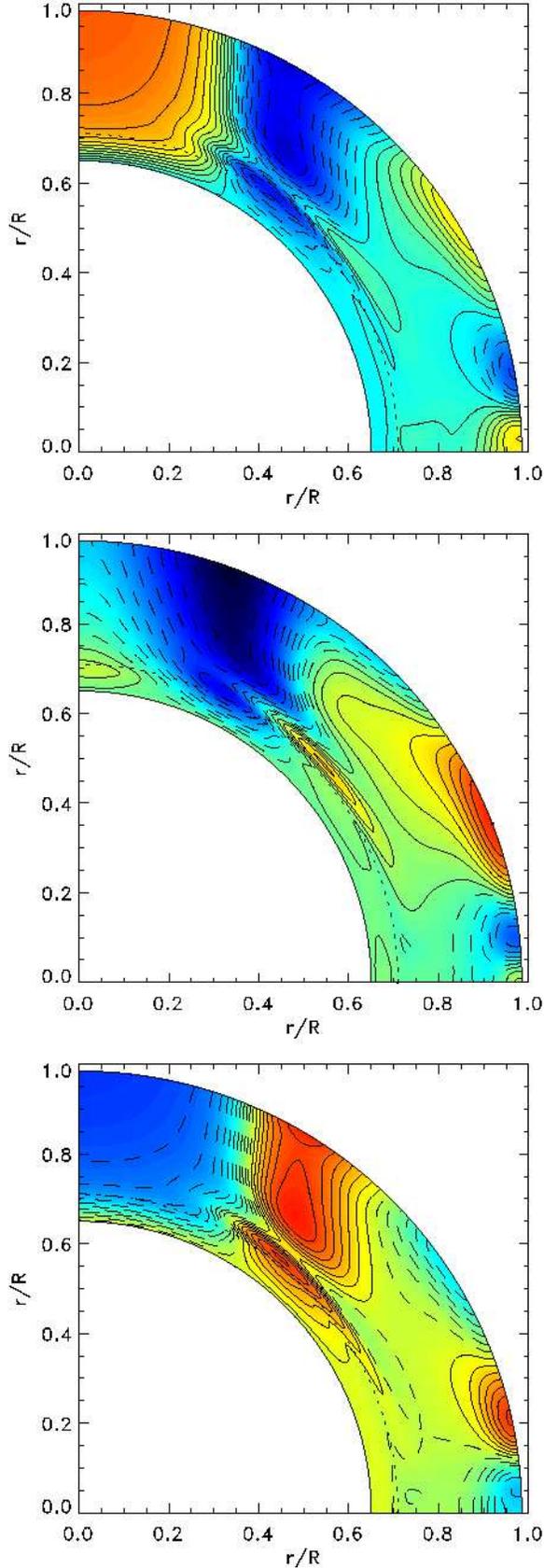}}
  \caption{Torsional oscillations for model with additional surface cooling.
    The frames (top to bottom) correspond to the times $t=1$, $t=3.5$, and 
    $t=6$, in Fig. \ref{f11}. While the polar branch of the torsional
    oscillations penetrates all the way to the base of the convection zone,
    the equatorial branch caused by surface cooling is confined close to
    the surface.
  }
  \label{f11}
\end{figure}

We can include the idea of Spruit in our model by parameterizing a surface
cooling term that depends on the toroidal field at the base of the
convection zone. To this end we change the upper boundary condition such
that the entropy gradient corresponds to an increase in energy flux.
With the diffusive convective energy flux 
$\vec{F}_c=\kappa_t\,\rh_0\,T_0\,c_v\vec{\nabla} s_1$ and 
$c_v=(\gamma-1)^{-1}p_0/(\rh_0\,T_0)$ this leads to
\begin{equation}
  \frac{\partial s_1}{\partial r}=-\frac{\gamma-1}{\kappa_t\,p_0}
  \,\epsilon F_{\odot}\,
  \left(\frac{(\sin\theta)^2\bar{B}_{\Phi, {\rm bc}}}{B_{\rm ref}}\right)^2\;.
\end{equation} 
Here $\bar{B}_{\Phi, {\rm bc}}(\theta)$ is the toroidal field averaged 
according to Eq. (\ref{B_bc}) between $0.71\,R_{\odot}$ and 
$0.76\,R_{\odot}$, which is also used for the
Babcock-Leighton $\alpha$-effect. $B_{\rm ref}$ is a reference field
strength used for normalization, $\epsilon$ determines the amplitude 
of the flux enhancement as fraction of the solar luminosity. In order
to illustrate this effect and allow a comparison to the observed
solar torsional oscillation we introduce a factor $(\sin\theta)^2$ 
in front of $\bar{B}_{\Phi}$ to correct the latitudinal field strength 
profile of the butterfly diagram accordingly (this is consistent
with the $(\sin\theta)^2$ factor introduced in the $\alpha$-effect for the
same reason). We use a dynamo model with magnetic buoyancy switched off
and a smaller value of $\alpha$ ($0.1\,\vunit$) than before in order to 
obtain an amplitude of the torsional
oscillations close to the observed one (around $4$ nHz at the pole and
$2$ nHz at $60\degr$ latitude. Fig. \ref{f10} (top panel) shows
the torsional oscillation and butterfly diagram for the dynamo solution not
considering the surface cooling. The middle panel shows the solution with
surface cooling considered using the parameters $\epsilon=2\times 10^{-2}$ 
and $B_{\rm ref}=1\,\mbox{T}$. The cooling of the active region belt drives
a equatorward propagating torsional oscillation with an amplitude of
around $1.5$ nHz, with the peak values at the edges of the active region belt.
The amplitude of the associated temperature perturbations (shown in the bottom 
panel) is around $0.2$ K. The peak cooling rate in our model is close to
$0.75\%$ of the solar energy flux, the surface intergrated luminosity variation
corresponds to around $0.23\%$ luminosity change throughout the cycle, about
a factor of $3$ larger than the observed one. We emphasize that this value
is imposed at $r=0.985\,R_{\odot}$ and does not necessarily resemble the
value required in a more realistic model extending all the way into the
photosphere and using a more sophisticated description of convection than the
diffusion approximation.
Fig. \ref{f11} shows three snapshots of the evolution of the
torsional oscillation in a $r$-$\theta$-plane. While the poleward propagating
branch (driven by Lorentz force feedback on differential rotation) is 
penetrating all the way down to the base of the convection zone, the 
equatorward propagating branch (driven through the surface cooling) is more 
concentrated toward the surface. As a side effect of the surface cooling the
model also shows close to the surface an inflow into the active region belt 
with a peak amplitude of around $5\,\vunit$. This value is in agreement with 
the theory of \citet{Spruit:2003}, where the meridional component is a 
consequence of an Ekman boundary layer at the solar surface. 
\citet{Komm:etal:1993,Komm:1994} derived an average inflow into the active 
region belt of around $5\,\vunit$ from Kitt Peak magnetograms. More recently
also helioseismology showed a mean inflow into the active region belt
with an amplitude from $2\,\vunit$ to $8\,\vunit$ 
\citep{Zhao:Kosovichev:2004}. 

\subsection{Forcing of torsional oscillations through quenching of viscosity
and heat conductivity}

\begin{figure}
  \resizebox{\hsize}{!}{\includegraphics{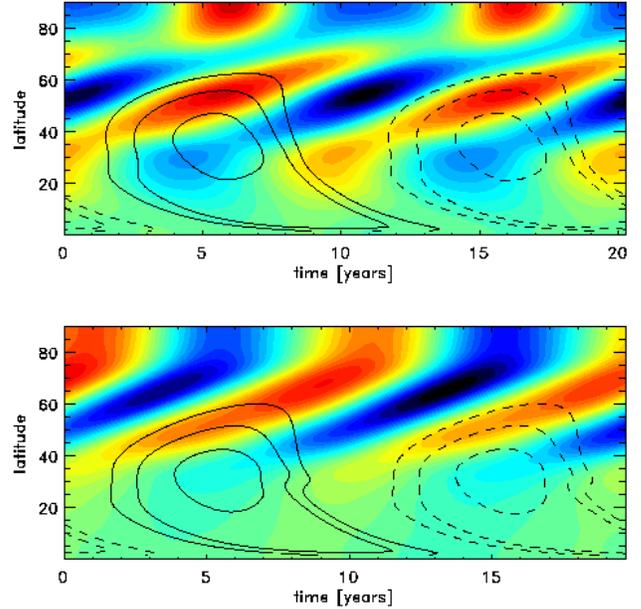}}
  \caption{Top: near surface torsional oscillation caused through quenching 
    of turbulent viscosity, the maximum amplitude is $0.77$ nHz; bottom: 
    torsional oscillation caused through quenching of turbulent viscosity 
    and heat conductivity, the maximum amplitude is $3.1$ nHz, respectively.
  }
  \label{f12}
\end{figure}

So far we only considered the 'macroscopic' Lorentz force feedback in terms
of the Lorentz force computed from the magnetic mean field. Alternatively
we can consider 'microscopic' feedback through quenching of turbulent 
motions. This type of feedback is typically used in kinematic models in terms
of $\alpha$-quenching to saturate the dynamo. We consider here the quenching 
of turbulent viscosity and heat conductivity. Since in our model the turbulent
viscosity scales the amplitude of the $\Lambda$-effect, quenching of $\nu_t$
reduces the energy input into the system and leads to a saturation of the
dynamo as consequence of a reduction in differential rotation similar to
the 'macroscopic' feedback discussed above. Additional quenching of turbulent
heat conductivity leads to changes in entropy profile of the convection zone.
Since the differential rotation is close to a baroclinic balance, a change
in the entropy profile also forces changes in the differential rotation.
Fig. \ref{f10} (top panel) shows torsional oscillations and the butterfly 
diagram obtained from a dynamo model with $\alpha_0=0.125\,\vunit$ and 
$\nu_t$ quenching with $B_{\rm eq}=1\,\mbox{T}$ according to Eq. 
(\ref{eta_quench}). Fig. \ref{f12} (bottom panel) shows torsional oscillations
resulting from additional quenching of the turbulent heat conductivity.
Note the different phase relation of the torsional oscillations with respect 
to the magnetic butterfly diagram compared to the results with 'macroscopic' 
Lorentz force feedback.
Quenching of turbulent heat conductivity alone is not a process that can 
efficiently saturate the dynamo, since it leads more to a modulation of
$\Omega$ rather than a reduction in the equator-pole difference. It can however
significantly change the torsional oscillation pattern. Since there are 
additional effects such as anisotropic heat transport, which are not considered
in our model, we emphasize here only the interesting result that
the amplitude of these thermally forced oscillations is quite significant.
Quenching of turbulent viscosity is along the lines of $\Lambda$-quenching
considered before by \citet{Kitchatinov:Pipin:1998,Kitchatinov:etal:1999,
Kueker:etal:1999}. 

\section{Parameter dependence}
\label{parameter}
The results presented in this paper were obtained using a mean field model
that requires parameterizations of unresolved processes. Therefore it is
necessary to test to which extent results are sensitive with respect to
details of the parameterizations used. 
Our differential rotation / meridional flow 
model has three important parameters: $n$ describing the profile of
the $\Lambda$-effect in latitude, $\Lambda_0$ determining the amplitude of the
$\Lambda$-effect, and the turbulent diffusivities $\nu_t$, $\kappa_t$ (both 
have same value in our model). $\Lambda_0$ has been chosen to get differential
rotation with the correct amplitude. For a fixed value of $\Lambda_0$,
$\nu_t$ and $\kappa_t$ determine the amplitude of the meridional flow. The
latter has been chosen to lead to dynamo simulations with a period close to
$22$ years.  The parameter $n$ determines the extent of the
meridional flow cell in latitude. Reasonable choices are in the range
$2-4$ (for values above $4$ differential rotation is confined to low 
latitudes, which also contradicts observations). Within this range the 
influence on the solutions is rather limited (see Fig. \ref{f9}). A fourth
parameter not discussed in this paper is the direction of the turbulent 
angular momentum flux with respect to the axis of rotation. In this paper
we used a fixed value of $\lambda=15\degr$, as has been used also for most
models in \citet{Rempel:2005}. Significantly larger values lead to more
complicated meridional flow patterns (permanent reverse cell in high 
latitudes),
smaller values require and increase of $\Lambda_0$ to maintain the differential
rotation amplitude, which in return leads to larger meridional flow speeds,
requiring significantly lower values for $\nu_t$ and $\kappa_t$ to obtain
a solar-like dynamo period. We computed solutions (not shown in this paper)
with the parameters $\lambda=7.5\degr$, $\Lambda_0=2$ and 
$\nu_t=\kappa_t=1.25\times 10^8\,\dunit$ that show no significant difference,
except that the amplitude of the torsional oscillations is around $50\%$
larger due to the reduced viscous damping. Therefore calibrating our 
reference model to be 
solar-like does not leave a lot of choice for the parameters of the
differential rotation model. Using such a calibrated reference model, the
following main results are not strongly dependent on the additional
dynamo parameters ($\alpha$-effect, $\eta_t$) providing the magnetic 
diffusivity is low enough to be in an advection dominated regime: 
The dynamo saturates with
a toroidal (mean)field strength of about $1.2$ to $1.5\,\mbox{T}$ at the base 
of the convection zone, the maximum magnetic flux produced during a cycle is
around $10^{16}\,\mbox{Wb}$, the energy converted on average into magnetic
energy is $0.1\%$ to $0.2\%$ of the solar luminosity (more likely $0.1$ 
incorporating additional constraints set by helioseismic observations of 
torsional oscillations). Another important result is that the dynamo saturates
already through reduction of differential rotation at a field strength
where the direct feedback on the meridional flow is insignificant. Therefore
the fundamental character of the flux-transport dynamo remains unaltered
(the non-linear feedback tends to concentrate magnetic activity even closer 
toward the equator).

\section{Implications for solar dynamo}
\label{discussion}
In this paper we discussed 'dynamic' flux-transport dynamos by combining
a mean field differential rotation model with a mean field dynamo model.
The main results from this study are the following
\begin{enumerate}
\item[$\bullet$] The non-kinematic Babcock-Leighton flux-transport dynamo
  saturates through a reduction of the amount of differential rotation at a 
  toroidal field strength of around $1.5$ T. This field strength is found to
  be fairly independent from particular choices of the strength of the
  $\alpha$-effect and value of magnetic diffusivity. The energy conversion 
  rate of the dynamo is around $0.001\,L_{\odot}$.
\item[$\bullet$] The Lorentz force feedback (at the saturation field 
  strength of around $1.5$ T) on the meridional flow is not strong enough to 
  switch off the equatorward transport of toroidal field required for the
  operation of the flux-transport dynamo. This is consistent with the
  findings of \citet{Rempel:2005:transport} who studied the modification of 
  meridional flow by an imposed stationary toroidal magnetic field. 
\item[$\bullet$] The Lorentz force feedback on differential rotation leads
  to torsional oscillations that are comparable (in terms of amplitude and
  phase with respect to the magnetic butterfly) to the high latitude branch 
  of the pattern inferred from helioseismology. The low latitude branch cannot
  be explained through Lorentz force feedback in our model. A model including
  thermal forcing through increased radiative losses in the active region belt
  \citep{Spruit:2003} also produces the low latitude branch. Thermal forcing
  is found to be very efficient in the sense that temperature variation of 
  only a few tenth of a degree are sufficient to explain the observed 
  amplitudes of torsional oscillations.
\end{enumerate}

The results presented here cannot answer the question of whether the solar
dynamo is a flux-transport dynamo or not; however, they make the case stronger
for a flux transport dynamo since they demonstrate that flux-transport dynamos
also function in the non-kinematic regime (the meridional flow is strong
enough to transport toroidal field of around $1.5$ T strength equatorward).
Additional to that the torsional oscillations caused by the macroscopic 
Lorentz force in a flux-transport dynamo are in agreement with helioseismic
results (in terms of amplitude and phase relation). With 'in agreement' we
mean here that there are no features produced that are not observed; however
there are observed features (the low latitude branch) that require additional
physics such as thermal forcing \citep{Spruit:2003}. The idea of thermal
forcing of the low latitude branch is also consistent with a mean inflow
into the active region belt of the order of $5\,\vunit$ that has been observed 
in magnetograms \citep{Komm:etal:1993,Komm:1994} and inferred from
helioseismology \citep{Zhao:Kosovichev:2004}. We emphasize that thermal
forcing is a very efficient process, since only tiny temperature fluctuations
of the order of a few tenth of a degree can drive large scale flows with the
observed amplitude. This opens the possibility that also other processes such
as the magnetic quenching of the convective energy flux in the solar convection
zone contribute.

\citet{Covas:etal:2000,Covas:etal:2004,Covas:etal:2005} presented
a $\alpha\Omega$-dynamo model (no meridional flow) including a simplified 
momentum equation for considering feedback on differential rotation. In their 
simulation they were able to reproduce additional to the polar branch also
the low latitude oscillations pattern. The main difference between their and 
our model is that we use a flux-transport dynamo with a non-local 
Babcock-Leighton $\alpha$-effect, while they use a classic 
$\alpha\Omega$-dynamo model with a (negative) local $\alpha$-effect in the
convection zone to obtain an equatorward propagating dynamo wave. The 
requirement to have a propagating dynamo wave leads to a fixed phase relation
between poloidal and toroidal field, which automatically leads to a 
Lorentz force pattern propagating with the field and producing a torsional
oscillation pattern associated with the magnetic field. This constraint is
relaxed in a flux-transport dynamo, where the propagation of the activity belt
is a pure advection effect. Given the fact that the origin of the low latitude
oscillation pattern in uncertain and might be entirely surface driven, it is
difficult to judge to which extent this discrepancy is of concern for 
different types of dynamo models and might rule out certain approaches.

We emphasize here that it is essential for models of torsional oscillations
to include the full momentum equation, since the Taylor-Proudman constraint
applies also to perturbations of $\Omega$ ($\partial \Omega_1/\partial z =0$)
and therefore significantly alters the phase relation of the oscillations.
Torsional oscillations with $\partial \Omega_1/\partial z \neq 0$
(as observed in the sun at low latitudes) require additional thermal 
perturbations, which makes also the consideration of an energy equation 
essential and favors explanations such as the one proposed by 
\citet{Spruit:2003}.

Torsional oscillations contain valuable information about the dynamo 
processes in the solar interior and are helpful to impose additional
constraints on dynamo models. Their interpretation and relation
to the dynamo generated magnetic field is complicated since different 
processes can produce torsional oscillations. As a first step toward using
torsional oscillations to probe solar cycle related processes in the solar 
interior it is important to test the sensitivity of helioseismic inversions
to dynamo generated rotation modulations in the solar interior,
especially close to the base of the convection zone. 
This work is currently done by \citet{Howe:etal:2004,Howe:etal:2006} for
different helioseismic techniques using artificial data and also model 
results discussed in this paper.

The model presented here sets strong constraints for the strength of toroidal
field at the base of the convection zone that can be achieved through the
shear by differential rotation. We find in our model a value of around 
$1.5$ T ($15$ kG) as upper limit. These results are in agreement with recent 
findings of \citet{Gilman:Rempel:2005} who showed that significantly larger 
field strength would require a very strong mechanism replenishing energy to 
differential rotation (replenishment time-scale of less than a month). 
The fairly low field strength (compared to convective equipartition) raises 
the question of how these fields can rise through the convection zone and 
form coherent sunspots at the surface. On the one hand recent work by Y. Fan 
(2006, private communication) suggests that 3D simulations of rising flux tubes
can reproduce most sunspot properties starting with field strengths around 
$2$ to $3$ T at the base of the convection zone (as opposed to thin flux tube 
simulations requiring around $10$ T); 
on the other hand recent work by \citet{Brummell:etal:2002:buo,
Cattaneo:etal:2006} is questioning the existence of coherent buoyant 
structures. In our model these uncertainties are hidden behind the 
parameterization of the Babcock-Leighton $\alpha$-effect. Since the existence 
of this effect is known from surface observations (it is the essential 
ingredient in models describing the evolution of the surface magnetic field
assimilating real observations \citep{Schrijver:etal:2002,Baumann:etal:2004,
Baumann:etal:2006,Wang:etal:2005}) this is of secondary concern
for the model presented here; it is however of primary interest for 
understanding the microphysics beyond the mean field approach.

\acknowledgements
The author thanks M. Dikpati, P.~A. Gilman and K.~B. MacGregor for very 
helpful comments on a draft of this paper. Very helpful suggestions by the
anonymous referee are also acknowledged.  

\bibliographystyle{natbib/apj}
\bibliography{natbib/apj-jour,natbib/papref}

\end{document}